\documentclass{osa-article}

\journal{oe}

\articletype{Research Article}

\usepackage{lineno}
\usepackage{subfigure}
\usepackage{caption}
\usepackage{comment}
\usepackage{bm} 
\usepackage{graphicx}
\usepackage{appendix}
\newcommand{\bfsfI}{\mbox{\sffamily\bfseries{I}}}
\newcommand{\bfsfM}{\mbox{\sffamily\bfseries{M}}}
\newcommand{\bfsfP}{\mbox{\sffamily\bfseries{P}}}
\newcommand{\bfsfS}{\mbox{\sffamily\bfseries{S}}}

\begin{document}
\title{Graphene multilayers  for coherent perfect absorption: effects of interlayer separation}

\author{Devashish Pandey,\authormark{1,*} Sanshui Xiao,\authormark{1,2,3} and Martijn Wubs\authormark{1,2,3}}

\address{\authormark{1}Department of Electrical and Photonics Engineering, Technical University of Denmark, DK-2800 Kgs. Lyngby, Denmark\\
\authormark{2}NanoPhoton - Center for Nanophotonics, Technical University of Denmark, Ørsteds Plads 345A, DK-2800 Kgs. Lyngby, Denmark\\
\authormark{3}Centre for Nanostructured Graphene, Technical University of Denmark,  DK-2800 Kgs. Lyngby, Denmark}

\email{\authormark{*}depan@.dtu.dk}

\begin{abstract}
We present a model study to estimate the sensitivity of the optical absorption of multilayered graphene structure to the subnanometer interlayer separation.
Starting from a transfer-matrix formalism we derive semi-analytical expressions for the far-field observables. Neglecting the interlayer separation, results in upper bounds to the absorption of 50\% for real-valued sheet conductivities, exactly the value needed for coherent perfect absorption (CPA), while for complex-valued conductivities we identify upper bounds that are always lower. For pristine graphene the number of layers required to attain this maximum is found to be fixed by the fine structure constant. For finite interlayer separations we find that this upper bound of absorption only exists until a particular value of interlayer separation ($D_{\rm lim}$) which is less than the realistic interlayer separation in graphene multilayers. Beyond this value, we find a strong dependence of absorption with the interlayer separation. For an infinite number of graphene layers a closed-form  analytical expression for the absorption is derived, based on a continued-fraction analysis that also leads to a simple expression for $D_{\rm lim}$. Our comparison with experiments  illustrates that multilayer Van der Waals crystals suitable for CPA can be more accurately modelled as electronically independent layers and more reliable predictions of their optical properties can be obtained if their subnanometer  interlayer separations are carefully accounted for.
\end{abstract}          

\section{Introduction and motivation}

Multilayer or stratified media have numerous uses in optics including so-called Bragg reflectors~\cite{saleh2019fundamentals}. Bragg mirrors are standard equipment in optical labs, and Bragg scattering has many modern optical applications~\cite{horsley2014revisiting}. The functionality of all-dielectric Bragg mirrors arises due to the multiple repetition of a few-layer unit cell, so that within a stop band of frequencies light can be almost totally reflected with low loss, even though a single unit cell reflects little. Bragg's law was originally formulated for X-ray diffraction off lattice planes in atomic crystals~~\cite{Bragg:1913a}. 

With the discovery of graphene, a single lattice plane,  as a stable form of carbon in 2004~\cite{novoselov2004electric}, the very active research field of two dimensional (2D) materials was born, and families of other stable single-atom thick sheets of materials were identified and studied theoretically. Many of these have been synthesized while many others still wait to be discovered ~\cite{haastrup2018computational,gjerding2021recent}. Novel optical properties offered by 2D materials include the following: Semi-metals such as graphene have voltage-tunable optical properties, unlike bulk noble metals~\cite{ju2011graphene}; the absorption of light in a monolayer of pristine graphene is 2.3$\%$, independent of frequency and proportional to the fine structure constant~\cite{nair2008fine}; unlike in bulk media, transition metal dichalcogenides have excitons that are stable at room temperature and interact strongly with light~\cite{gonccalves2020strong}; monolayers of transition metal dichalcogenides have direct band gaps whereas two or more layers including bulk material have indirect band gaps~\cite{RevModPhys.90.021001}. Finally, the large-band gap material  hexagonal boron nitride is widely used as a spacer layer, but also can host bright color centers even at room temperature~\cite{tran2016quantum,fischer2021controlled}.

Layers of different 2D materials can be combined to form multilayers on the atomic scale with new functionalities, known as Van der Waals materials \cite{novoselov20162d}. For example, 2D heterostructures can host voltage-tunable interlayer excitons. When combining 2D monolayers and Van der Waals materials with the ''classical'' optical multilayers,  the myriad of applications of stratified media increases even further. For example, graphene-based multilayer structures can act as tunable hyperbolic metamaterials~\cite{iorsh2013hyperbolic,othman2013graphene}. 

Here our main focus is not so much on the 2D monolayers, but rather on the interlayer spacing between them. For example, in graphite the separation between the graphene sheets is $0.334~\mbox{nm}$ ~\cite{Cakmak:2019a}, while layer-by-layer grown multilayer graphene (MLG) has an interlayer spacing in the range $0.55-0.7$ nm~\cite{wu2009synthesis}. We investigate in which situations these different separations on the atomic scale will affect the optical properties of the multilayers. 
It is however already clear that the interlayer spacing in some cases can be neglected. For example, in the famous work by Nair \textit{et al}~\cite{nair2008fine} it was found that the transmittance of few-layer graphene depends only on the product of the fine structure constant and the number of layers, and the interlayer spacing could be left out of the model. 

We will consider multilayer absorption in particular. Infinitely thin sheets have a theoretically maximal single-port absorption of 50\%~\cite{thongrattanasiri2012complete,baranov2017coherent}. 
There are many works that report an enhanced absorption in graphene for a wide frequency range as reviewed in Ref.~\cite{guo2018graphene}, while some even confirm close to 100\% absorption in monolayer graphene by designing ingenious complex photonic environments for infrared light~\cite{thongrattanasiri2012complete,graphene_monolayer_CPA}. 
Intriguingly, lossy beam splitters with a 50\% single-port absorption can lead to 100\% absorption when illuminating two ports. Indeed, two coherent  beams with identical amplitudes and phases incident on such a beam splitter will lead to complete destructive interference of the reflected and transmitted waves, leading to 100\% absorption. This is referred to as coherent perfect absorption (CPA), an intriguing mechanism to control light with light in linear optics~\cite{Chong:2010a,PhysRevA.83.055804,pirruccio2013coherent, hardal2019quantum,vetlugin2021coherent,CPA_degenerate_OC}. Multilayer graphene has already been used for this purpose~\cite{rao2014coherent, roger2015coherent,zanotto2017coherent}. Our focus here is with which multilayer graphene structures one can expect to achieve a  single-port optical absorption of 50\%. For such graphene-based CPA beam splitters it is less clear and has not been addressed whether the interlayer spacing can be neglected when modelling their absorption properties. This is a main motivation for the present study. 

The structure of our paper is as follows: In Section~\ref{Sec:basics} we introduce the basics in our notation, including the transfer-matrix description of monolayers,  the absorption in multilayer structure when neglecting interlayer separation, and the corresponding fundamental absorption limits of graphene. Section~\ref{Sec:absorption} then discusses multilayer absorption when taking finite interlayer separation into account, particularly for pristine graphene. In Section~\ref{Sec:graphite_Delf} we compare our model with experimental results. We end with our conclusions and outlook in Section~\ref{Sec:conclusions}.  

\section{Neglecting interlayer separation: mono- and multilayers, absorption bounds}\label{Sec:basics}

A main goal of this study is to find out what is the effect of interlayer separation on the optical properties of multilayers. Here we first neglect these interlayer separation and obtain analytical expressions for transmission, reflection, and absorption of multilayers, as well as fundamental limits to absorption by these multilayers. Later on we will compare these analytical results with numerical ones for finite interlayer separation.

\subsection{Monolayers: 2D and 3D conductivities and transfer matrices}\label{Sec:monolayers}

\begin{figure}
\centering
\includegraphics[width=0.7\textwidth]{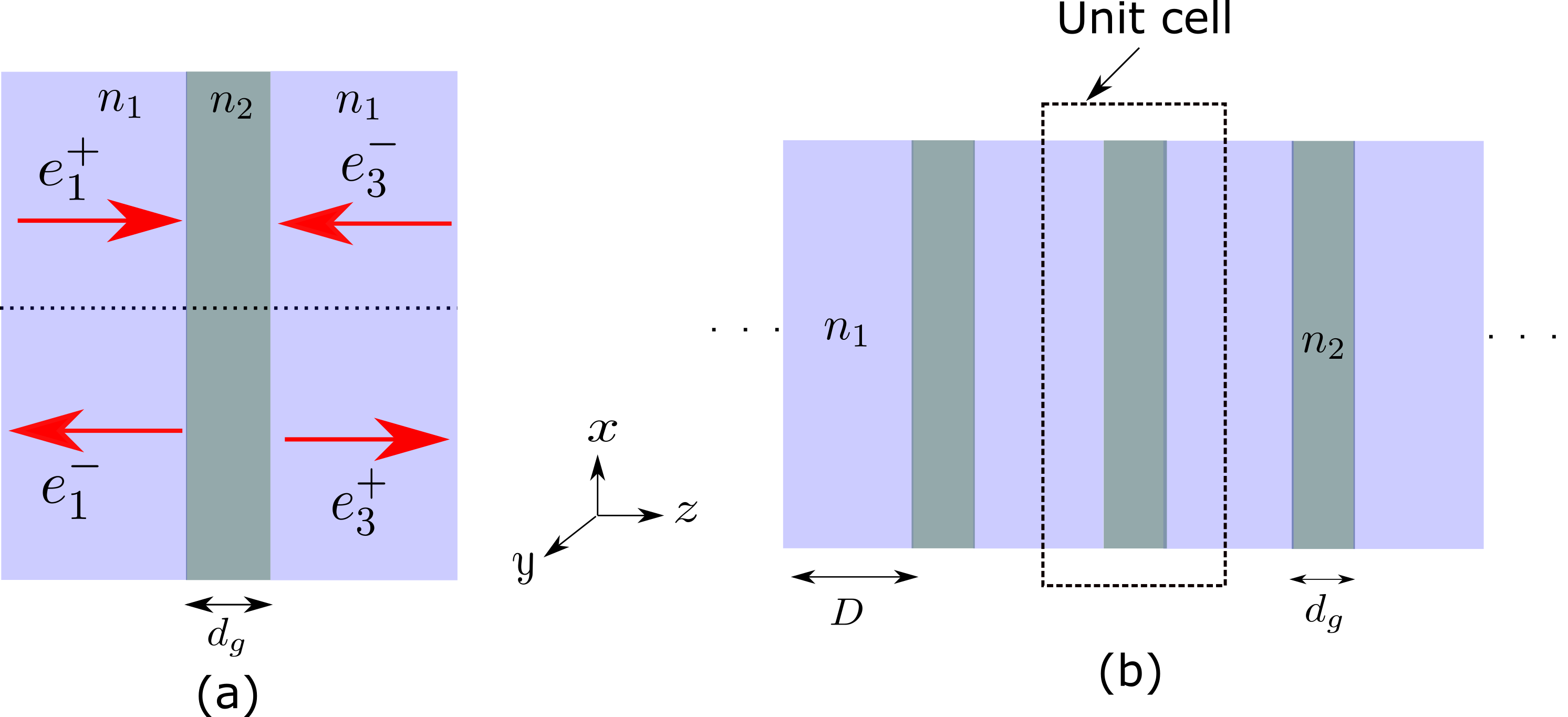}
\caption{(a) A 2D monolayer shown as gray slab of thickness $d_g$ and with refractive index $n_2$ with the input (output) field amplitudes $e^+_1$, $e^-_1$ ($e^+_3$, $e^-_3$) (b) A multilayer structure with 2D monolayers (shown in gray) separated by the interlayer dielectric of thickness $D$ and refractive index $n_1$. The unit cell is denoted by the dotted black box.  }
\label{graphene_layers}
\end{figure}

Monolayers are typically characterized by their two-dimensional surface conductivity $\sigma_{2D}({\bf k}_\parallel, \omega)$ exhibiting both frequency- and in-plane wavevector dispersion. 
Monolayers can be modellled as being infinitely thin and described by the three-dimensional conductivity  $\sigma_{3D}({\bf k}_\parallel, \omega) =\sigma_{2D}({\bf k}_\parallel, \omega)\delta(z)$, where $z$ is the coordinate perpendicular to the in-plane direction. But it can also be useful to think of the monolayer as a slab instead, centered around $z=0$ with sub-nanometer finite thickness $d_g$ that by construction has the same spatially-integrated conductivity, so $\sigma_{3D}({\bf k}_\parallel, \omega)=\sigma_{2D}({\bf k}_\parallel, \omega)/d_g$ within the slab and zero outside. One can then understand the transmission, reflection, and absorption in the monolayer from the well-known transfer matrix of a slab~\cite{saleh2019fundamentals}, which is the combination of two interfaces and homogeneous propagation in between, and where the relative dielectric function inside the monolayer slab is given by $\varepsilon_r = 1 + i   \sigma_{2D}/\varepsilon_0\omega d_g$ as shown in Fig.~\ref{graphene_layers}(a) with $n_2=\sqrt{\varepsilon_r}$. This dielectric function describes a lossy monolayer, unless $\sigma_{2D}$ is purely imaginary. Even though atoms have finite sizes and monolayers have finite thicknesses, for the optical properties it is very accurate to take the $d_g\to 0$ limit in the slab transfer matrix. By taking this limit and considering normal incidence (${\bf k}_{\parallel} = {\bm 0}$), we find the transfer and scattering matrices
\begin{equation}
\bfsfM = \left( \begin{array}{cc}
1 - \beta & - \beta \\
\beta & 1 + \beta \end{array}\right), \qquad
\bfsfS = \frac{1}{1+\beta}\left( \begin{array}{cc}
1  & - \beta \\
- \beta & 1  \end{array}\right),
\label{M_g}
\end{equation}
where $\beta\equiv \sigma_{2D}/(2\varepsilon_0 c)$ and  $\sigma_{2D} = \sigma_{2D}({\bf k}_\parallel = {\bm 0}, \omega)$. Because it gave this result in a simple way, it paid off to first treat a monolayer as a finite slab and then take the infinitely-thin limit again. 

The transfer matrix~(\ref{M_g}) is finite and does not depend on $d_g$ anymore. In principle it is possible to add a correction to first order in $d_g$ to Eq.~(\ref{M_g}), but we will not pursue that here. From the transfer matrix one obtains in the usual way the transmission amplitude $t=1/(1+\beta)$ and reflection amplitude $r=-\beta/(1+\beta)$~\cite{baranov2017coherent,Goncalves:2016a}, and hence transmission $T=|t|^2$, reflection $R= |r|^2$, and absorption $A=1-R-T$. The transfer matrix $\bfsfM$ in general represents a lossy medium since $1/(1+\beta)\neq 1/(1-\beta)^* $, while the equality holds for lossless reciprocal media, giving $A=0$.

\subsection{Multilayers: limits to absorption when neglecting interlayer separation}\label{Sec:arbitrarymultilayers}
After providing a suitable description of a monolayer we can proceed to model the multilayered structure. Various approaches  have been used in this regard, and focusing on graphene we can mention for example large scale tight-binding calculations~\cite{zhu2014optical}, scattering matrix formalisms~\cite{baranov2017coherent} and transfer matrix approaches~\cite{zhan2013transfer}. We use the latter and probe the sub-nanometer variation of the interlayer separation in the far field.  
Now we consider a multilayer structure with the same assumption of $d_g \to 0$ limit for the monolayer. This assumption implies that the influence of a monolayer on light can be described by a zero-thickness conducting sheet that has the same spatially integrated conductivity as that of a sheet of finite thickness $d_g$. When describing multilayers, one can then still assume that the monolayer sheets have a finite separation $D$ and the multilayer has a finite total thickness that is a multiple of $D$; in the limit $d_g\to 0$, the separation $D$ has become the period of the multilayer. It is often assumed that as far as optical properties are concerned, also $D$ can be taken to vanish in a transfer-matrix description. Here we briefly consider consequences of that assumption, before studying the effect of finite $D$ as our main objective later.

$N$ monolayers each with a sheet conductivity $\sigma_{2D}$ separated by zero separation are optically equivalent to a single monolayer with sheet conductivity $N \sigma_{2D}$.  (For more microscopic considerations of multilayer conductivity of graphene we refer to Ref.~\cite{min2009origin}.)  The total transfer matrix $\bfsfM_{\rm tot}$ is then as $\bfsfM$ in Eq.~(\ref{M_g}) but with $\beta$ replaced by $N \beta$. This result is also immediately found by calculating $\bfsfM_{\rm tot}$ as $\bfsfM^N$.    
The transmission, reflection and absorption by the multilayer are then given by~\cite{min2009origin,kuzmenko2008universal} 
\begin{equation}
    T_N=\frac{1}{|1+N\beta|^2},\;\quad R_N= |N\beta|^2 T_N,\; \quad A_N= 1 - R_N - T_N, \; 
    \label{T_R_A_zeroD}
\end{equation}
where the conductivity and hence also $\beta$ in general are complex-valued.  Clearly, in general the optical properties depend in a nonlinear fashion on $N$. Taking $N=1$ gives back the results for the monolayer. In the limit $|N \beta| \ll 1$ the multilayer interacts weakly with light and to first order we find 
\begin{equation}\label{T_R_A_zeroDlinearinN}
T_N = 1- 2N \mbox{Re}(\beta),\quad R_N = 0, \quad A_N = 2N \mbox{Re}(\beta) \qquad\mbox{(first order in $N$)}.
\end{equation}
In this limit the effects of the multilayer on light indeed vary linearly with the number of layers and reflections are negligible.

We are more interested in stronger light-matter interactions, especially in the question what is the maximum absorption that one could hope to get when varying the number of layers, for a given sheet conductivity. 
To that end we  rewrite $N \beta $ in polar representation as $p \mbox{e}^{i \theta}$, with $p,\theta$ both real-valued, so that $N$ only affects $p$ while $\theta$ depends only on $\sigma_{2D}$, independently of $N$. Then we can express $A_N$ in Eq.~\eqref{T_R_A_zeroD} in terms of $p$ and $\theta$ as 
\begin{equation}
A_N = \frac{2 p \cos \theta}{1 + p^2 + 2 p\cos \theta}\label{Eq:A_N_ptheta}
\end{equation}
For the situation $\theta = \pi/2$ we find that the absorption $A_N$ vanishes, which makes sense because  the 2D conductivity is then purely imaginary, corresponding to a real-valued dielectric function for the multilayer. The absorption in Eq.~(\ref{Eq:A_N_ptheta}) could even get negative if the cosine is negative, describing gain in the medium, which we do not consider here.
Now for a given frequency of light, $\sigma_{2D}$ is fixed so the value of $\theta$ is also fixed, but we can vary the number of layers to get maximum absorption. In our parametrization we look for the maximum of $A_N$ upon variation of $p$, while keeping $\theta$ constant. We arrive at the condition $p=1$, and it is interesting that this does not depend on $\theta$. For that optimal value of $p$ we thus find the condition on the number  of layers $N_{\rm max}$ that gives the maximum absorption
\begin{equation}
N_{\rm max} = |\beta|^{-1} = \frac{2 \varepsilon_0 c}{|\sigma_{2D}|} \qquad  \Rightarrow \qquad A_{N}^{\rm max} = \frac{\cos \theta}{1 + \cos \theta}.
\label{Eq:A_max}
\end{equation}
We find that $0 \leq A_N \leq 0.5$, with  0.5 being the fundamental upper limit of absorption for the special situation of a real-valued sheet conductivity ($\cos \theta=1$). For other derivations of this fundamental absorption limit of 50\% by zero-thickness sheets, see Refs.~\cite{thongrattanasiri2012complete,baranov2017coherent}. There the argument given for this limit is attributed to the negligible phase change of the incident radiation after scattering with the boundary of the subwavelength structure. In such circumstances the relation between reflection and transmission can be shown to be $t=1\pm r$ with `+' and `-' for s- and p-polarized waves respectively, which results in an absorption $A=1-|r|^2-|1\pm r|^2$ having a maximum value of 0.5~\cite{thongrattanasiri2012complete}. 

For complex-valued 2D conductivities $\sigma_{2D}$ this maximum of 50\% absorption cannot be reached by varying the number of layers. It is interesting that for complex conductivities precise upper bounds for absorption, all with  values below 0.5, are given by Eq.~(\ref{Eq:A_max}),  and that they only depend  on the  angle $\theta$ of the polar representation of $\sigma_{2D}$ in the complex plane. 
 
The above results make it challenging to make a 50\% absorbing beam splitter (a "CPA beam splitter"~\cite{hardal2019quantum}) using a 2D multilayer, because it is related to a fundamental upper limit of absorption: only if the sheet conductivity is purely real-valued can one find 50\% absorption (and nothing more) by optimizing the number of layers. For complex-valued sheet conductivities no such 50\% absorbing beam splitter can be made, at least not as long as the thickness of the multilayer can be neglected (which we do here). Before considering finite interlayer separations, let us first apply our above general results for arbitrary sheet conductivities and vanishing interlayer separation to graphene. 

\subsection{Example:  mono- and multilayers of pristine graphene}\label{Sec:graphenepristine}

There are numerous experiments determining the thickness of a single graphene layer which ranges from $0.1-1.7$ nm depending on the preparation methods, substrates and techniques~\cite{shearer2016accurate}. To understand its optical properties,  graphene has been modelled as a dielectric slab of constant thickness  treating the monolayer as a homogeneous medium with an effective thickness given by the interlayer spacing of the respective bulk material~\cite{li2014measurement,benameur2011visibility,simon2020single}. Alternatively it has been modeled using the surface current model~\cite{stauber2008optical,merano2016fresnel,zhan2013transfer} where graphene is defined as a sheet of infinitesimally small thickness. Arguments why the sheet conductivity  model would be superior can be found in Ref.~\cite{merano2016fresnel}, whereas the superiority of the slab model is claimed in Ref.~\cite{simon2020single}. Here we describe instead graphene as we have done above for arbitrary monolayers: we describe graphene as a slab but then determine the transfer matrix of graphene in the limit of vanishing thickness of this slab. The important thing in relation to the above slab/sheet discussion is that we distinguish the thickness $d_g$ of the monolayer slabs from the interlayer spacing that we denote by $D$, such that $d_g$ can be taken to zero while $D$ stays finite, similar to how Bragg scattering of x-rays by atomic planes is described~\cite{Bragg:1913a}.

It is well known that in the local random phase approximation (RPA) limit for zero temperature  the 2D conductivity of pristine graphene has a remarkably simple form, namely $\sigma_{2D} =\sigma_0= e^{^2}/4 \hbar$~\cite{nair2008fine,koppens2011graphene,bludov2013primer,luo2013plasmons}, and that this can be written in terms of the fine structure constant  $\alpha \equiv e^{2}/(4 \pi \varepsilon_{0}\hbar c)\simeq 1/137$, as  $\sigma_{2D}= \sigma_0 \equiv \varepsilon_{0} \pi c \alpha$. Our results of Sec.~\ref{Sec:monolayers} for arbitrary monolayers thus apply to pristine graphene when we substitute $\beta= \pi \alpha/2$, and similarly for our results for arbitrary multilayers in Sec.~\ref{Sec:arbitrarymultilayers}. 
For example, as long as $N \beta  \ll 1$ or $N \ll 2\times 137/\pi$, the absorption in a graphene multilayer grows approximately linearly with the number of layers and Eq.~(\ref{T_R_A_zeroDlinearinN}) now gives $A_N = N \pi \alpha$, which is the well-known 2.3\% absorption for every added layer~\cite{nair2008fine}. This linear regime brings us up to a total absorption of 20\% corresponding to $N\approx 9$ which is well in the regime $N \beta  \ll 1$. Since the conductivity of pristine graphene is real-valued, the maximum absorption of a multilayer given by Eq.~(\ref{Eq:A_max}) can be 0.5, provided that the number of layers is chosen as 
\begin{equation}
N_{\rm max} = 1/\beta = \frac{2 }{\pi \alpha} =  87,  \label{Eq:N87} 
\end{equation}
rounded off to an integer, which is an interesting number as it only depends on the fine structure constant. It agrees fairly well with Ref.~\cite{zanotto2017coherent} where a CPA regime is identified for 100 layers of graphene on top of a substrate, and where vanishing interlayer separations were assumed  just as we do until now. However, when taking finite interlayer separation into account below, we will find from our numerical investigations that the analysis leading to Eq.~(\ref{Eq:N87}) is too simplified and fewer layers are required to obtain 50\% absorption.  

We have neglected interlayer interactions, but in Refs.~\cite{zhu2014optical} and~\cite{carr2020mapping} these interactions were taken into account in tight-binding simulations for multilayer graphene. An analytical fitting function relating the transmission and number of graphene layers was proposed in Ref.~\cite{zhu2014optical} that is based on Eq.~(\ref{T_R_A_zeroD}) for non-interacting layers with negligible interlayer separation, namely $T(\omega)={1}/{[1+f(\omega)N\beta]^{2}}$. An experimental fit gave $f(\omega)=1.13$ instead of unity for $\lambda=550$ nm~\cite{zhu2014optical}. This modification is small, which makes sense because the neglected  Van der Waals interactions between layers are weak. 

\subsection{Example:  mono- and multilayers of doped graphene}\label{Sec:graphenedoped}

For doped graphene the conductivity can be dominated by a Drude conductivity~\cite{bludov2013primer}
$\sigma_{2D} = \nu /(\gamma - i \omega)$ where the parameter $\nu$ depends on the Fermi energy as $\nu = \sigma_0 (4 E_F / (\pi \hbar))$. This conductivity is defined in the regime $\hbar \omega\ll E_F$ where the absorption results due to the intraband transitions. Unlike for pristine graphene, this conductivity is complex-valued, so by Eq.~(\ref{Eq:A_max}) the multi-layer absorption in the $D \to 0$ limit will be lower than 0.5. Analogous to pristine graphene in Sec.~\ref{Sec:graphenepristine}, we can now  find what is the maximum absorption of a doped multilayer there after optimizing the number of layers (and still assuming that interlayer separation can be neglected). The polar decomposition of the conductivity reads $\sigma_{2D} = \nu /(\gamma - i \omega) = |\sigma_{2D}| \mbox{e}^{i \theta}$, so that we can identify $|\sigma_{2D}| = \gamma /\sqrt{\gamma^2 + \omega^2}$ and $\theta = \tanh{(\omega/\gamma)}$. Using this in~(\ref{Eq:A_max}) for the Drude model gives the maximal absorption, as long as the interlayer separation $D$ can be neglected, 
\begin{equation}
N_{\rm max} = \frac{2\varepsilon_0 c}{\nu}(\omega^2 + \gamma^2)^{1/2}\qquad \Rightarrow \qquad A_{N}^{\rm max} = \frac{\gamma}{\gamma + (\omega^2+\gamma^2 )^{1/2}} \quad \mbox{(for Drude model)}
\end{equation}
We find that  this maximal absorption is independent of the Fermi energy $E_F$. For optical frequencies, which have our main interest,  we have $\omega \gg \gamma$ so that $A_{N}^{\rm max} \ll 1$. By contrast,  in the static limit $\omega \ll \gamma$, we find that large absorption is possible again, and indeed with the maximal  value of 0.5 in that limit, in agreement with the discussion of almost-real conductivities in Ref.~\cite{zanotto2017coherent}. 
 
\section{Interlayer-dependent absorption of multilayer graphene}\label{Sec:absorption}
While previously we calculated properties of multilayer graphene in the $D\rightarrow 0$ limit, now we will take into account the finite separation $D$ between the layers. We will limit our discussion to that of the pristine graphene hereafter for which $\beta = \pi\alpha/2$, and use  incident light of wavelength $\lambda = 550$ nm in all our numerical results. Our analysis simplifies by defining a unit cell as shown in Fig.~\ref{graphene_layers}(b), with corresponding  unit cell transfer matrix $\bfsfM_U=\bfsfP\bfsfM\bfsfP$. Here, $\bfsfM$ is the transfer matrix of a graphene monolayer as in Eq.~(\ref{M_g}), while  $\bfsfP$ is the propagation matrix in air of thickness $D/2$, given by a diagonal matrix with $e^{\pm i\phi/2}$ on the two diagonals, where $\phi \equiv k_0 D$. This gives the unit cell transfer matrix
\begin{equation}
    \bfsfM_U=\left( \begin{array}{cc} 
    (1 - \beta) e^{i \phi} & - \beta \\ \beta & (1+ \beta) e^{-i \phi} \end{array}\right).
    \label{unit_cell2}
\end{equation}

The total transfer matrix for a layered structure with $N$ such unit cells is then given by the $N^{\rm th}$ power of this matrix. This can be simplified using the Chebyshev identity which has been used extensively to describe light in lossless periodic systems~\cite{yeh1977electromagnetic,saleh2019fundamentals} and has been extended to lossy systems as well~\cite{horsley2014revisiting, iorsh2013hyperbolic, smirnova2014multilayer, nefedov2013perfect}. The identity only holds for unimodular matrices. i.e. $\mbox{Det}(M_U)=1$, which indeed applies to Eq.~\eqref{unit_cell2} even in the presence of loss. Therefore using the identity we have $(\bfsfM_U)^N = U_N \bfsfM_U-U_{N-1} \bfsfI$, 
where $U_N=\sin (N \Psi)/\sin \Psi$ are the Chebyshev coefficients of the second kind and are a function of $\cos \Psi$, with $U_0=0, U_1=1$ while the  the  coefficients for $N\ge 2$ follow the recursive relation  $U_N(\cos\Psi)=U_2 U_{N-1}(\cos\Psi)-U_{N-2  }(\cos\Psi)$. Here the complex Bloch phase in general is defined by $\Psi=\cos^{-1}\left[(a+d)/{2}\right]$, where $a$ and $d$ are the diagonal elements of the unit cell matrix $\bfsfM_U$. It can be a source of confusion that only for lossless structures this complex Bloch phase $\Psi$ can be found from the identity $\cos \Psi=\mbox{Re}(1/t_U)$, with $t_U$ defined as the transmission coefficient of the unit cell~\cite{yeh1977electromagnetic,saleh2019fundamentals}. Using instead the definition of $\Psi$ in combination with Eq.~\eqref{unit_cell2}, we find the Bloch phase for our 2D multilayer as  $\Psi=\cos^{-1}\big(\cos\phi-i\beta\sin\phi\big)$. By Taylor approximations we find $\Psi\approx \big({\phi^2+2i\beta\phi}\big)^{1/2}$, which is an excellent approximation within the range of values of $N$ and $D$ that we will use. 
\begin{figure}
\centering
\subfigure{\label{trans_undoped}\includegraphics[width=0.42\textwidth, height=40mm]{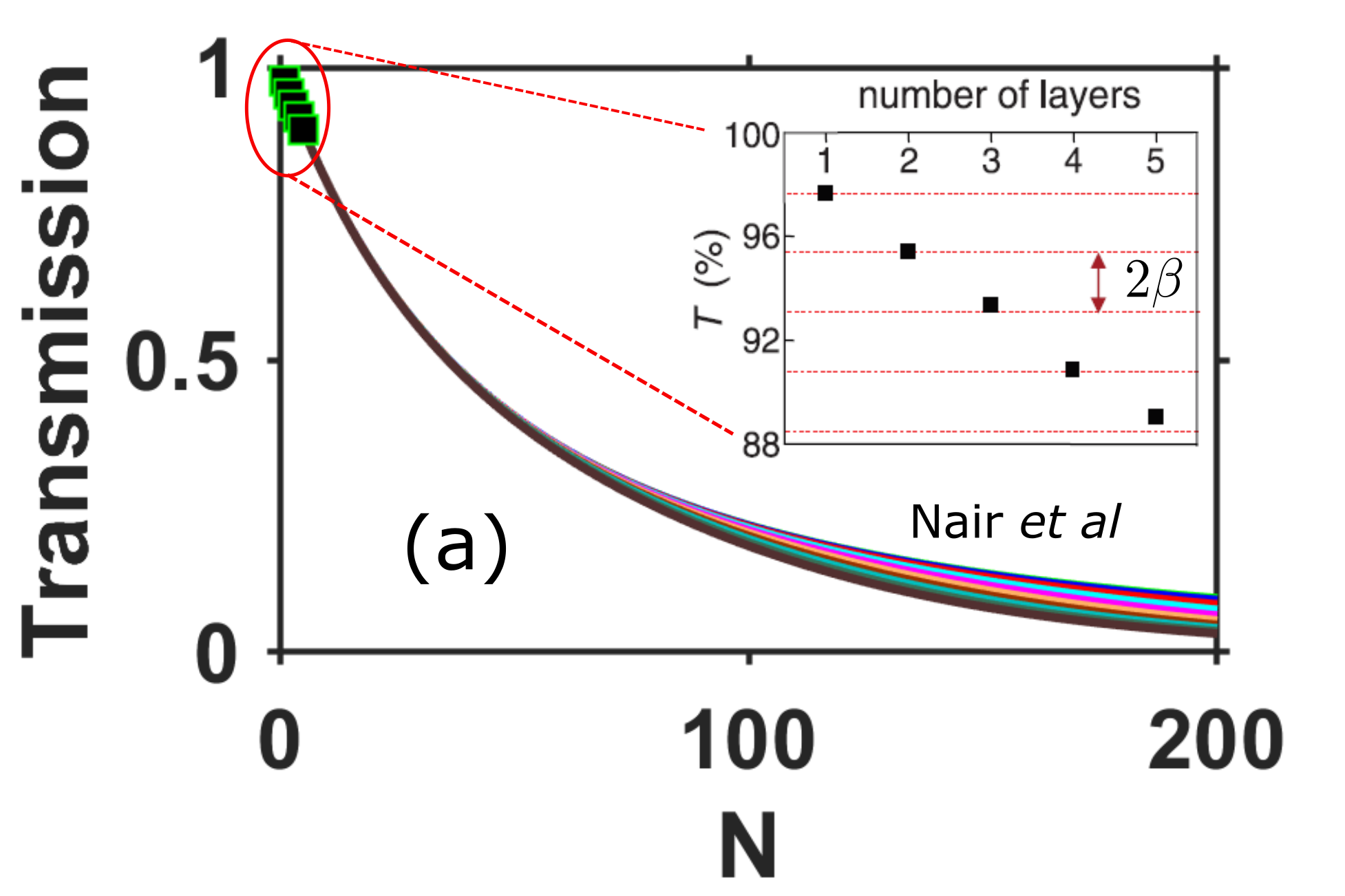}}
\subfigure{\label{absorp_undoped}\includegraphics[width=0.50\textwidth, height=40mm]{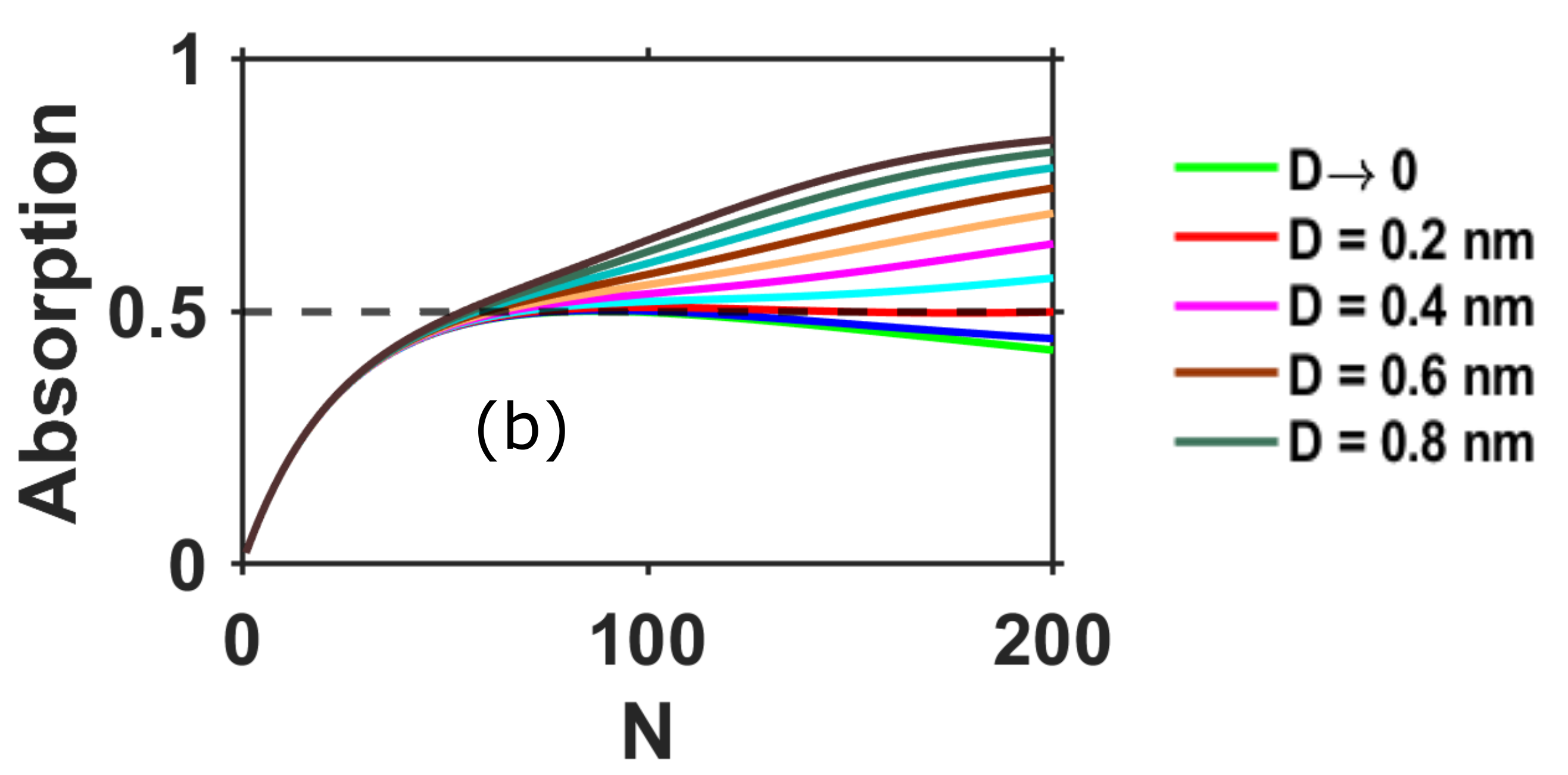}}
\caption{(a)~Transmission through a multilayer of pristine graphene as a function of the number of layers. Graphs correspond to different fixed values of the interlayer separation $D$, with only half of them color-coded alternately to the right of panel~(b).  Inset: experimental linear dependence of the transmission with a small number of graphene layers, from Ref.~\cite{nair2008fine}.  (b)~Corresponding graphs for multilayer absorption. The black dashed line $A_N=0.5$ separates graphs with a maximum absorption at finite $N$ from  those with an ever growing $A_N$.  }\label{Fig:TA_N_finiteD}
\end{figure}
We can now use the Chebyshev identity to obtain the intensity transmission, reflection, and absorption coefficients of the multilayer as
\begin{equation}\label{Eq:TRA_finiteD}
    T_N=\frac{1}{|U_N(1+\beta)e^{-i\phi}-U_{N-1}|^2},\;\quad R_N=|\beta U_N|^2 T_N, \qquad A_N = 1 - R_N - T_N,
    \end{equation}
which agrees with Eq.~(\ref{T_R_A_zeroD}) in the limit of vanishing interlayer separation, as it should. 

In Fig.~\ref{Fig:TA_N_finiteD} we depict the multilayer transmission and absorption of Eq.~(\ref{Eq:TRA_finiteD}) as a function of $N$, for several fixed  values of the interlayer separation $D$. Not all of these values for $D$ are realistic, so part of  Fig.~\ref{Fig:TA_N_finiteD} should in the first place be seen as a numerical experiment, while realistic values for $D$ are discussed later. From the transmission in panel~(a) it is clear that only for a few layers of graphene does the transmission agrees with the classic experiment in Ref.~\cite{nair2008fine}, irrespective of the chosen interlayer separation $D$, while for a hundred layers or more, the dependence of the transmission on $D$ is still modest. By contrast,   multilayer  absorption  in panel~(b) depends much more sensitively on the interlayer separation. For $D=0$ (light green curve) the maximum absorption is 50\% at $N=87$, as derived in Eq.~(\ref{Eq:N87}). Interestingly we see two families of plots separated by the line $A_N = 0.5$, where the curves below this line have their maxima all at $A_N \approx 0.5$, while for the other family the absorption increases monotonically with $N$.   

\begin{figure}
    \centering
    \includegraphics[width=0.8\textwidth]{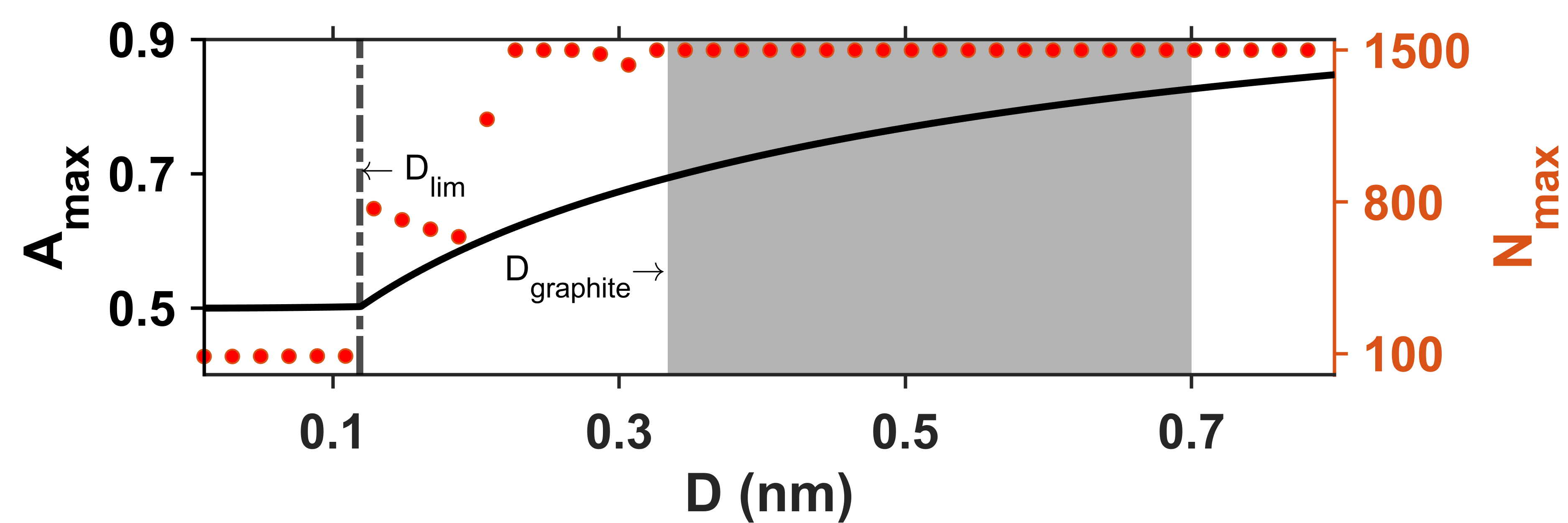}
    \caption{Maximum absorption (black solid line) of light in multilayered graphene and its corresponding number of layers ($N_{\rm max}$) (shown by red square symbols) versus the interlayer separation $D$. The value $D_{\rm lim}$, indicated by the dash-dotted line,  is the largest value of $D$ for which the absorption peaks at $A_N \simeq 0.5$ for a certain finite number of layers. For $D > D_{\rm lim}$, the absorption increases monotonically with $D$ and $N$. Filled gray area: the range of realistic values of the interlayer separation.  }
    \label{A_max}
\end{figure}

To further clarify this interesting behavior, we determined the maximal absorption in every graph of Fig.~\ref{Fig:TA_N_finiteD}(b)  which we plot in Fig.~\ref{A_max} as a function of the corresponding interlayer separation. In doing so we chose even higher values of $N = 1500$ beyond which the magnitude of the absorption was seen to become constant. It can be seen that until an interlayer separation of $D_{\rm lim}=0.12\;\mbox{nm}$, denoted by a vertical dashed black line in Fig.~\ref{A_max}, there exists a maximum absorption that is more or less constant at $50\%$. Also the corresponding number of layers in this range of $D$ lies within 100 layers as shown by the red square symbols in Fig.~\ref{A_max}, in line with Ref.~\cite{zanotto2017coherent}. However, in the regime $D>D_{\rm lim}$, i.e. to the right of the vertical dashed line in Fig.~\ref{A_max},  the maximum absorption is no longer bounded by $50\%$.

As to the question what interlayer separation can be considered realistic,  in graphite the interlayer separation is $0.334~\mbox{nm}$, a value that can be tuned towards larger values~\cite{Cakmak:2019a}.  When multilayer graphene is produced layer by layer, then interlayer separations will typically be slightly larger and experimental interlayer separations in the range  $0.55$ nm-$0.7$ nm  are given in~\cite{wu2009synthesis}. The gray shaded area in Fig.~\ref{A_max} corresponds to realistic values of $D$ of $0.334~\mbox{nm}$  to $0.7$~nm. It follows from Fig.~\ref{A_max} that for realistic values of $D$ the bound  $A^{\rm max}_N=0.5$ does not apply and multilayer absorption increases beyond 50\% for a fixed value of $N$, as the  interlayer separation is increased. Therefore  50-percent-absorbing beam splitters made of multilayer graphene always need to be described taking their finite interlayer separation into account.  The precise number of layers $N_{50 \%}$  that will result in the 50\% absorption depends sensitively on the value of the interlayer separation $D$, as  will be investigated numerically below. So in summary, for the region in Fig.~\ref{A_max} to the left of $D_{\rm lim}$ we can very well approximate  the multilayer transmission, reflection and absorption by our analytical results for $D =0$ in Eq.~\eqref{T_R_A_zeroD}. However, to the right of $D_{\rm lim}$ this approximation fails and we need to take the finite value of $D$ into account and use Eqs.~\eqref{Eq:TRA_finiteD} instead. This is due to a non-negligible phase change of the scattered light contrary to the situation discussed in Sec.~\ref{Sec:arbitrarymultilayers}. Therefore the maximum absorption can exceed $50\%$, even for strongly subwavelength structures, as illustrated in Figs.~\ref{Fig:TA_N_finiteD}(b) and~\ref{A_max}. 

 \subsection{Deducing the number of layers leading to $50\%$ absorption}

\begin{figure}
    \centering
    \includegraphics[width=0.7\textwidth]{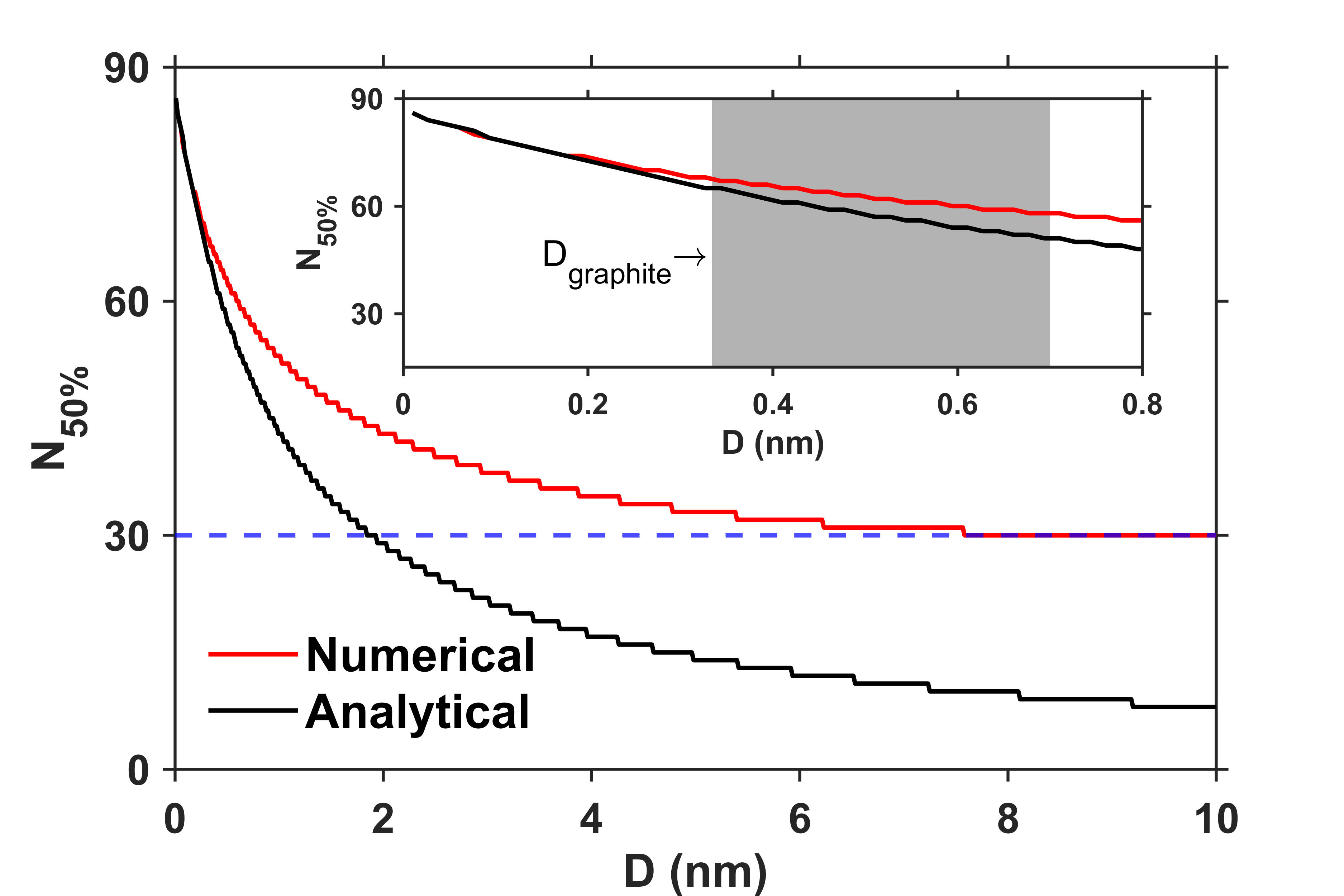}
    \caption{Number of layers of  pristine graphene that will lead to a $50\%$ absorption ($N_{50\%}$),  versus the interlayer separation between the individual layers. The numerical results (obtained from Eq.~\eqref{A_50}) and the analytical approximation Eq.~\eqref{N_50_ana} agree reasonably well for realistic values of $D$ (see gray area in the inset). }
    \label{N_50_ana_num}
\end{figure}

By imposing $50\%$ absorption condition in Eq.~\eqref{Eq:TRA_finiteD} we obtain the condition
\begin{equation}
     \frac{1}{2}=\frac{1+|\beta U_N|^2}{|U_N(1+\beta)e^{-i\phi}-U_{N-1}|^2}\Bigg|_{N=N_{50\%}}.
     \label{A_50}
\end{equation}
This equation is then solved numerically and the red solid lines in Fig.~\ref{N_50_ana_num} 
 provide insight which combinations of interlayer separation and number of pristine graphene layers will give the sought $50\%$ absorption ($N_{50\%}$). In the limit $D \to 0$, we find back $N_{50\%}=87$ as derived previously in Eq.~(\ref{Eq:N87}). As the interlayer separation is increased, fewer layers are needed to get 50\% absorption. For example, for the measured value of $D= 0.6~\mbox{nm}$~\cite{wu2009synthesis} we find $N_{50\%}= 60$. For the minimal realistic value  $D= D_{\rm graphite}= 0.334~\mbox{nm}$ the corresponding number is 67  layers. This makes it very clear that, from the perspective of our model, interlayer separation should be taken into account when determining the required number of layers to get 50\% absorption. This  is surprising since  67 layers of graphene in graphite are only 22 nm thick, less than five percent of a wavelength at optical frequencies. And for realistic interlayer separation  in the range $0.334-0.7~\mbox{nm}$, the required number of layers to get 50\% absorption will be at least 58, as can be seen from the red curve at $D = 0.7~\mbox{nm}$ in the inset of Fig.~\ref{N_50_ana_num}. This is still considerably larger than the experimental value of 30 layers in Refs.~\cite{rao2014coherent,roger2015coherent}. This difference can partly be attributed to the large theoretical uncertainty in $N_{50\%}$ in the almost horizontal plateaus in the in absorption curves near $A_N = 0.5$ upon variation of $N$, as we saw in Fig.~\ref{Fig:TA_N_finiteD}(b). For example, for $D=0.334$ nm, $N$ varies in the range from $33$ to $67$ as the absorption increases from  $40\%$ to $50\%$.
 For interlayer separations that are an order or magnitude larger than what could be considered realistic, i.e. for $D \ge 5~\mbox{nm}$, Fig.~\ref{N_50_ana_num} illustrates that $N_{50\%}$ would drop to $30$.
 
An analytical estimate for $N_{50\%}$ can be found by making  the first-order Taylor expansion $U_N \approx N$ in Eq.~\eqref{A_50}, giving
\begin{equation}
    N_{50\%}=\frac{(1+\beta)\phi^2/2+\phi-\beta}{(1+\beta)\phi^2-\beta^2}.
    \label{N_50_ana}
\end{equation}
This approximation becomes exact in the limit $D \to 0$, and proves to be a reasonably good approximation for realistic values of $D$, as illustrated in Fig.~\ref{N_50_ana_num}.

\subsection{Limiting case of semi-infinite multilayers ($N\rightarrow \infty$)}
\begin{figure}
    \centering
    \includegraphics[width=0.9\textwidth]{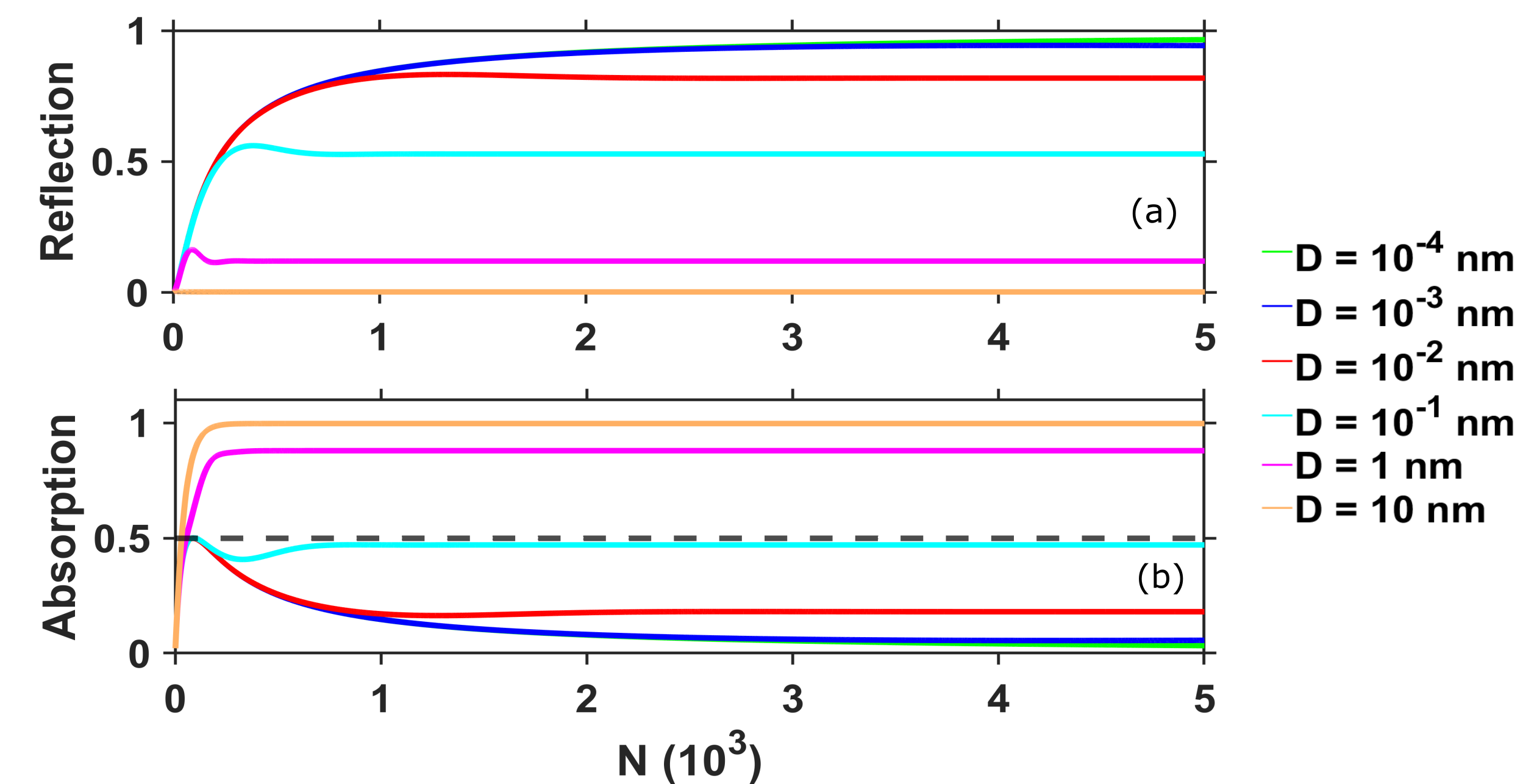}
    \caption{(a) Reflection with respect to the number of layers $N$ , up til values $N\gg 1/\beta$, for various fixed values of $D$. Based on Eq.~\eqref{Eq:TRA_finiteD}. (b) Corresponding absorption graphs. The dashed black line corresponds to 50\% absorption.}
   \label{r_t_infinite}
\end{figure}

%
Next we use Eq.~\eqref{Eq:TRA_finiteD} to show in Fig.~\ref{r_t_infinite} the dependence of the reflection and absorption on the number of layers, especially the asymptotic absorption for a large number of layers ($N\gg 1/\beta$), a limit that was not yet reached in Fig.~\ref{Fig:TA_N_finiteD}(b). In Fig.~\ref{r_t_infinite}, the two values  of 0.1 and 1.0~nm for $D$ are close to realistic, while the other values merely give an idea of the sensitivity of the results on the interlayer separation. 
Again as in Fig.~\ref{Fig:TA_N_finiteD}(b) before, in Fig.~\ref{r_t_infinite} we see a clear bifurcation of the family of absorption plots, where in one family (with $D \le D_{\rm lim}$) the  absorption peaks close to 50\%, while in the other family (with $D > D_{\rm lim}$) the absorption increases monotonically with $D$. We now also can see that the reflection and absorption values have converged as a function of $N$ in the regime $N\gg 1/\beta$. (This was also the reason to choose $N=1500\gg 1/\beta$ in Fig.~\ref{A_max} to assure convergence of absorption.) With this numerical intuition at hand we can now derive analytical expressions for these asymptotic values, based on a continued-fraction analysis: from Eq.~\eqref{Eq:TRA_finiteD}, we find the amplitude reflection  $r=\beta /[{(1+\beta)e^{-i\phi}-U_{N-1}/U_N}]$. Now defining $\zeta=U_{N}/U_{N-1}$ we get a continued fraction due to the recursive relations of the Chebyshev coefficient of the form $\zeta=U_2-U_{N-2}/U_{N-1}$. Since we are interested in the limit $N\rightarrow \infty$ the above relation will have infinite continued fractions, and for the reflection of the semi-infinite slab we obtain
\begin{eqnarray}\label{Eq:semiinfiniteR}
    R_\infty&=&\frac{\beta^2 }{|(1+\beta)e^{-i\phi}-e^{ i\Psi}|^2}.
    \label{R_infty}
 \end{eqnarray}
\begin{figure}
    \centering
    \includegraphics[width=0.8\textwidth]{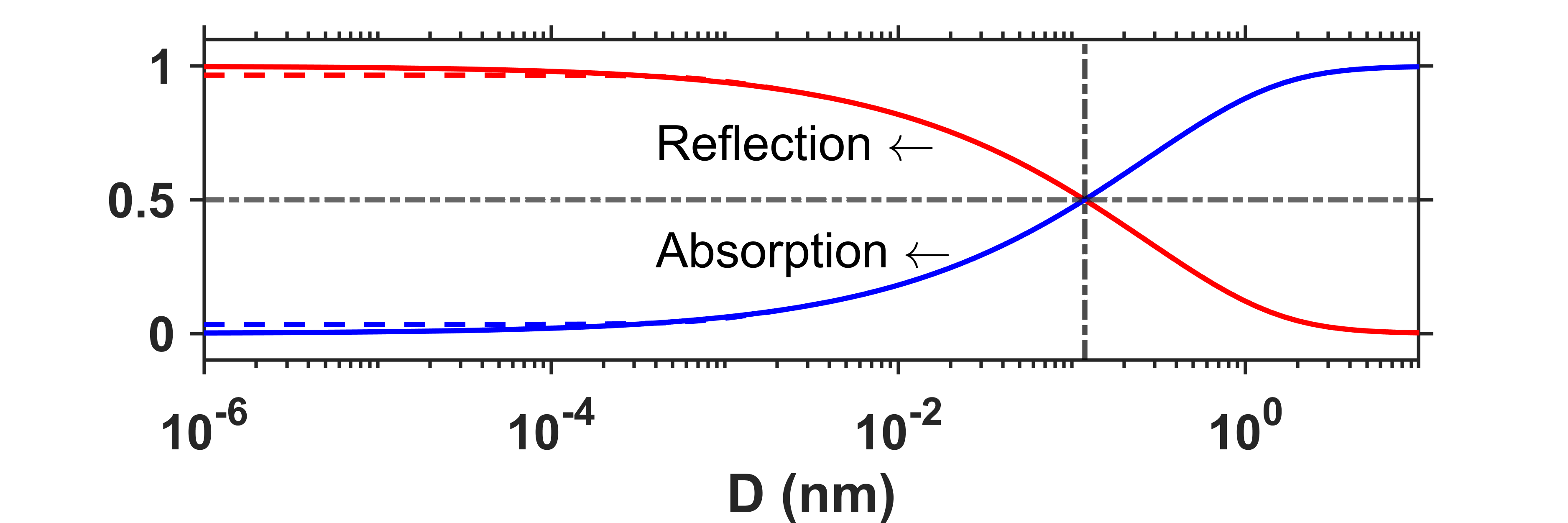}
    \caption{Asymptotic reflection (absorption) in red (blue) solid lines as function of  interlayer separation, in the limit  $N\rightarrow \infty$, as given by Eq.~\eqref{R_infty}.  The corresponding graphs for  finite $N=5000$, based on Eq.~\eqref{Eq:TRA_finiteD}, are shown in dashed lines.}
    \label{r_t_infity_d}
\end{figure}
In Fig.~\ref{r_t_infity_d}  we use this result to depict $A_\infty$ ($R_\infty$) and find that the asymptotic value of absorption (reflection) increases (decreases) with  interlayer separation. Moreover, the numerical computation using Eq.~\eqref{Eq:TRA_finiteD} for $N=5000$ layers (for which $N\gg 1/\beta$) can indeed be seen to converge to the analytical expression in Eq.~(\ref{Eq:semiinfiniteR}) for $N\rightarrow \infty$ for realistic values of $D$.
Thus in the range of realistic values of $D$, an MLG structure with $N\gg 1/\beta$ is a good absorber and a bad reflector, unlike in the expressions in Eq.~(\ref{T_R_A_zeroDlinearinN}) that were linearized in $N$, and which  of course are not valid in the limit $N\to \infty$. From Eq.~\eqref{R_infty} we can also deduce the value of $D_{\rm lim}$, which is the value of the interlayer separation that gives a maximum absorption of $50\%$. Defining $e^{-i\phi}\approx 1-i\phi$ and $\Psi\approx\sqrt{\beta\phi}(1+i)/\sqrt{2}$, we can arrive at an equation of the form $2\sqrt{\beta}(1+\beta)\phi^{3/2}+2\beta\sqrt{\beta}\phi^{1/2}+2\beta\phi-\beta^2=0$ which when solved leads to a very simple analytical form 
\begin{equation}\label{Eq:Dlim}
D_{\rm lim} \approx \frac{\beta}{9k_0}=\frac{\pi\alpha}{18 k_0}.
\end{equation}
For $\lambda = 550$ nm we find $D_{\rm lim} = 0.11$ nm, in good agreement with our numerical simulation in Fig.~\ref{r_t_infity_d}. 
For graphite we can turn the question around: given its fixed  interlayer separation  $D=0.334$ nm, what would be the wavelength for which there is a maximal absorption of 50\% for a specific finite thickness? Using $D_{\rm lim} = 0.334$ in Eq.~(\ref{Eq:Dlim})  we find $\lambda \approx 1650$ nm as the wavelength at which the absorption in graphite will peak at $50\%$. The corresponding thickness of graphite  can be extracted numerically similar to Fig.~\ref{r_t_infinite}(b),  but now for $\lambda = 1650$ nm and $D = 0.334$ nm, and comes out to be 270 nm.

\section{Comparison with graphite and experimental validation}
\label{Sec:graphite_Delf}
\begin{figure}
     \centering
     \includegraphics[width=0.75\textwidth]{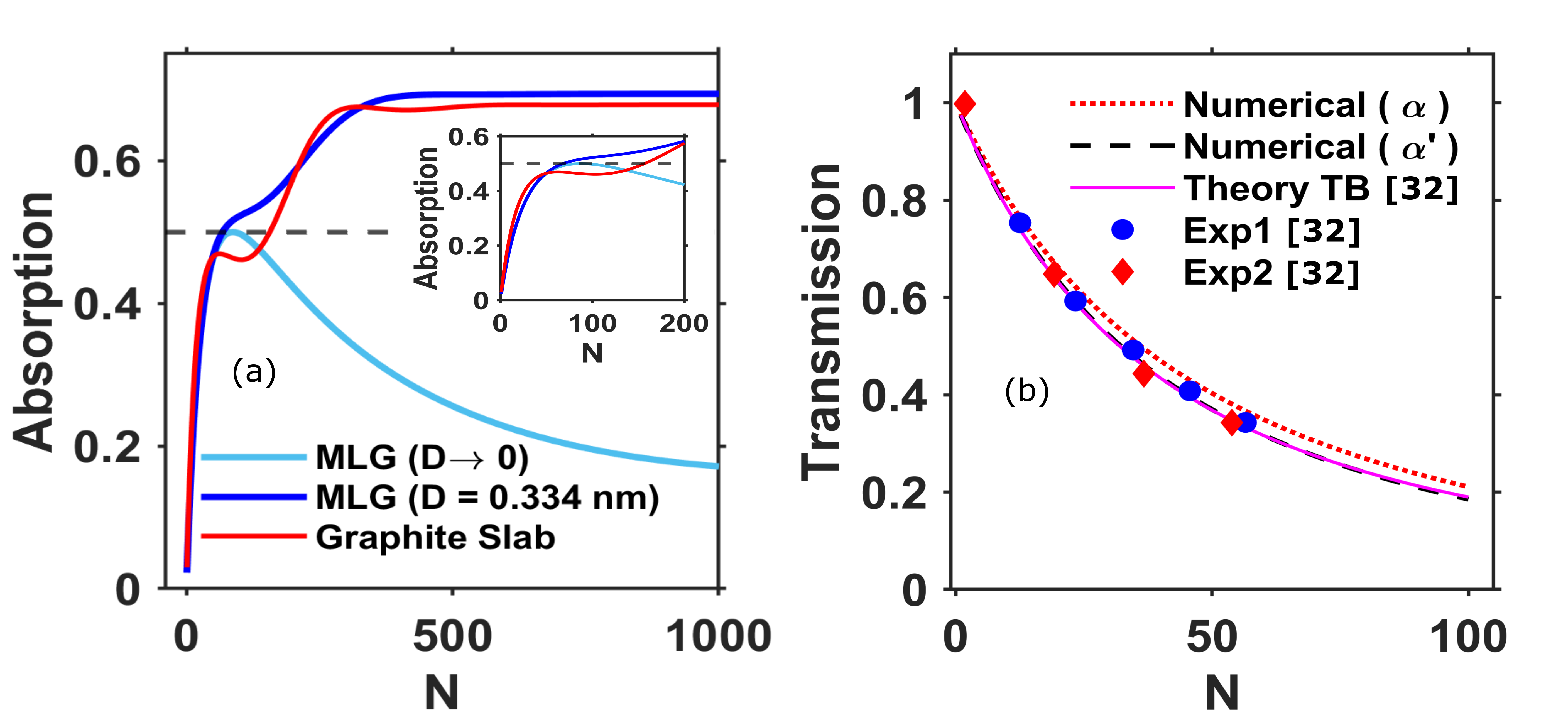}
     \caption{(a) Absorption in MLG with $D \rightarrow 0$ (shown in light blue solid line), for $D = 0.334$ nm (shown in dark blue solid line), and for a graphite slab with thickness = $0.334\times N$ nm (in red solid line) with the complex refractive index $n$ versus the number of layers. The black dashed line corresponds to a $50\%$ absorption. Inset: zoomed image of the figure for values of $N$ until 200. (b) Comparison of our numerical simulations (red dotted line) with numerically corrected  $\alpha'$ (black dashed line) derived from panel~(a) both using Eq.~\eqref{Eq:TRA_finiteD} compared with the tight-binding theory (magenta solid line) and experimental results (blue circles and red diamond symbols) reproduced from Ref.~\cite{zhu2014optical}.}
     \label{graphite_graphene_delf}
 \end{figure}
  In this section we will compare our model describing the  graphene $N$-layer structure ($D=0.334$ nm) with a graphite slab of thickness $0.334\times N$ that we describe by an experimentally obtained bulk complex refractive index $n=a+ib$. In Fig.~\ref{graphite_graphene_delf}(a) we see that  the two corresponding absorption curves agree fairly well. However,  the largest deviation occurs in our region of interest around $A_N = 0.5$, making CPA beam splitters based on graphene challenging to model in a simple way.  The  deviations between the two curves can be attributed to the interlayer interactions that we have neglected in our multilayer approach, or to the fact that we described nanoslabs of graphite by bulk parameters, or both. Our present study does not tell which of the two descriptions is the more accurate around $A_N = 0.5$.
If we however assume that the bulk description of graphite is accurate around $A_N = 0.5$, then we could fit the properties of a graphene layer to enforce agreement of its multilayers with graphite. We describe this pragmatic approach and then discuss its use.  

 From the refractive index of graphite, which for $\lambda=550$ nm is $n=2.7164+1.4848i$~\cite{djurivsic1999optical}, we can deduce the 2D conductivity as $\sigma_{2D}'=\pi\alpha'\varepsilon_0c=2ab\omega\varepsilon_0d_g$. We can now use $\sigma_{2D}'$ in our MLG simulation instead of $\sigma_{2D}=\pi\alpha\varepsilon_0c$, or equivalently $\alpha'$ instead of the fine structure constant $\alpha$. The $\alpha'$ has thereby become a fitting parameter that also incorporates the correction due to the interlayer interactions, and is given by $\alpha' = (4ab/\lambda)d_g$. Now tuning $d_g$ in such a way that the graphite slab model and the MLG model (with $\alpha'$ incorporated) show maximal agreement, we find $d_g \approx 0.278$ nm and the corresponding $\alpha' = 1.12\alpha$. It is interesting to see that our numerically extracted $d_g$ is within the reasonable range of graphene thickness as measured in Ref.~\cite{shearer2016accurate}. Using this $\alpha'$ instead of $\alpha$ in our calculation of transmission in Eq.~\eqref{Eq:TRA_finiteD}, we find very good agreement both with the experiments and with the tight-binding simulations of Ref.~\cite{zhu2014optical}, as shown in Fig.~\ref{graphite_graphene_delf}(b). Neglecting the interlayer separation (light blue curve in Fig.~\ref{graphite_graphene_delf}(a)), results in an error that becomes quite pronounced beyond the regime $N>1/\beta$.
 
\section{Conclusions}\label{Sec:conclusions}

We have studied multilayer graphene in the parameter region of interest for coherent perfect absorption, close to 50\% absorption. First we neglected interlayer separations and found a maximally possible value of the multilayer absorption that is given by Eq.~(\ref{Eq:A_max}), which never reaches to 50\% irrespective of the number of layers, unless the single-sheet conductivities are real-valued. When taking finite interlayer separations into account, multilayer transmission curves did not change appreciably, but the absorption curves did: two families of curves were seen in Fig.~\ref{Fig:TA_N_finiteD}(b), one family with a maximum absorption around 50\% for a finite number of layers, and the other family with an absorption that keeps growing with the number of layers. We also identified the limiting value of interlayer separation that separates the two families of absorption curves quantified by $D_{\rm lim}$ as shown in Fig.~\ref{A_max}. Realistic values for interlayer separation turn out to be larger than this limiting value, so that more accurate values for absorption are obtained by taking these interlayer separations into account, even for multilayers with subwavelength thicknesses.

We used a transfer-matrix approach where interlayer separation could be incorporated, but (electronic) interlayer interactions strictly speaking could not.  The advantage of the transfer-matrix approach is that we could readily obtain new analytical formulae for the maximal aborption in Eq.~(\ref{Eq:A_max}) for negligible interlayer separation and for arbitrary 2D conductivities. Also for finite $D$ could analytical estimates be found:  Chebyshev identities helped to find the number of layers that gives 50\% absorption in Eq.~(\ref{N_50_ana}) and the reflection of infinitely many layers in Eq.~(\ref{Eq:semiinfiniteR}), the latter based on a continued-fraction analysis. This analysis also helped in determining the limiting interlayer separation ($D_{\rm lim}$) of Eq.~(\ref{Eq:Dlim}).

A basic assumption of the transfer-matrix approach remains that the layers are electronically independent.
However, with a phenomenological  procedure where the fine structure constant in the graphene conductivity became a fitting parameter, we could find very good agreement with both experiments and with  tight-binding calculations of Ref.~\cite{zhu2014optical} by an increase of the fine structure constant of only 12\%. This illustrates that the neglected interlayer Van der Waals interactions indeed are weak, and gives an estimate for the accuracy of our analysis.

We focused here on multilayer graphene surrounded by air, and more general configurations of graphene and/or other 2D Van der Waals materials encapsulated in different dielectric environments can be modelled analogously.  Our results illustrate that multilayer Van der Waals crystals suitable for CPA can be more accurately modelled as electronically independent layers and more reliable predictions of their optical properties can be obtained if their subnanometer  interlayer separations are carefully accounted for.

\begin{backmatter}
\bmsection{Funding}
M.W. and S. X. acknowledge the support by the Danish National Research Foundation through NanoPhoton Center for Nanophotonics, Grant No. DNRF147, and Center for Nanostructured Graphene, Grant No. DNRF103. M.W. and D.P acknowledge the Independent Research Fund Denmark Natural Sciences (Project No. 0135-00403B).  S.X. acknowledges the Independent Research Fund Denmark (Project No. 9041-00333B, 2032-00351B).
\bmsection{Acknowledgments}
D.P. acknowledges Mads A. Jørgensen for stimulating discussions.
\bmsection{Disclosures}
The authors declare no conflicts of interest.
\bmsection{Data availability}
Data and code used in this paper are not publicly available but can be obtained from authors upon request.

\end{backmatter}
\bibliography{ref}
\end{document}